\documentclass[conference]{IEEEtran}
\IEEEoverridecommandlockouts

\usepackage[boxed, linesnumbered, ruled]{algorithm2e}
\usepackage{cite}
\usepackage{amsmath,amssymb,amsfonts, amsthm}
\usepackage{hyperref}
\usepackage{booktabs}
\usepackage{graphicx}
\usepackage{listings}
\usepackage{lstlangarm}
\usepackage{subcaption}
\usepackage{textcomp}
\usepackage{xcolor}
\usepackage{wasysym}

\def\BibTeX{{\rm B\kern-.05em{\sc i\kern-.025em b}\kern-.08em
    T\kern-.1667em\lower.7ex\hbox{E}\kern-.125emX}}

\definecolor{cerulean}{HTML}{006BA4}
\definecolor{mortargrey}{HTML}{595959}
\definecolor{pumpkin}{HTML}{FF800E}
\definecolor{brickred}{HTML}{CB4154}

\title{Double Strike: Breaking Approximation-Based Side-Channel Countermeasures for DNNs}
\author{
\IEEEauthorblockN{
Lorenzo Casalino\IEEEauthorrefmark{1},
Maria Méndez Real\IEEEauthorrefmark{2},
Jean-Christophe Prévotet\IEEEauthorrefmark{3},
Rubén Salvador\IEEEauthorrefmark{1}
}

\IEEEauthorblockA{\IEEEauthorrefmark{1} CentraleSupélec, Inria, IRISA, CNRS -- \{lorenzo.casalino, ruben.salvador\}@inria.fr}
\IEEEauthorblockA{\IEEEauthorrefmark{2} Lab-STICC, Univ. Bretagne-Sud, UMR 6285 -- maria.mendez-real@univ-ubs.fr}
\IEEEauthorblockA{\IEEEauthorrefmark{3} Univ. Rennes, INSA Rennes, CNRS, IETR-UMR 6164 -- jean-christophe.prevotet@insa-rennes.fr}}

\newcommand{\repourl}{\url{https://doi.org/10.5281/zenodo.17969535}}
\newcommand{\dataseturl}{\url{https://doi.org/10.5281/zenodo.17965887}}

\newcommand{\myvec}[1]{\textbf{#1}}

\newcommand{\change}[1]{\textcolor{brickred}{#1}}
\newcommand{\func}[1]{\texttt{#1}}

\newcommand{\acro}[1]{\textsc{#1}}
\newcommand{\arm}{\acro{arm}}
\newcommand{\axc}{\acro{AxC}}
\newcommand{\cmfour}{\acro{cortex-m4}}
\newcommand{\cnn}{\acro{cnn}}
\newcommand{\cwlite}{\acro{cwlite}}
\newcommand{\cpa}{\acro{cpa}}
\newcommand{\cpu}{\acro{cpu}}
\newcommand{\dnn}{\acro{dnn}}
\newcommand{\fpga}{\acro{fpga}}
\newcommand{\hw}{\acro{hw}}
\newcommand{\iapam}{\acro{}{IaPAM}}
\newcommand{\ip}{\acro{ip}}

\newcommand{\gpu}{\acro{gpu}}
\newcommand{\mac}{\acro{mac}}
\newcommand{\macpruning}{\acro{MACPruning}}
\newcommand{\guessentr}{\acro{ge}}
\newcommand{\mlp}{\acro{mlp}}

\newcommand{\sca}{\acro{sca}}
\newcommand{\thumbtwo}{\acro{thumb-2}}

\begin{document}

\maketitle

\begin{abstract}
Deep neural networks (\dnn{}s), which support services such as driving assistants and medical diagnoses, undergo lengthy and expensive training procedures. Therefore, the training's outcome – the \dnn{} weights -- represents a significant intellectual property asset to protect.
Side-channel analysis (\sca{}) has recently appeared as an effective approach to recover this confidential asset from \dnn{} implementations.
In response, researchers have proposed to defend \dnn{} implementations through classic side-channel countermeasures, at the cost of higher energy consumption, inference time, and resource utilisation.
Following a different approach, Ding et al. (HOST'25) introduced \macpruning{}, a novel \sca{} countermeasure based on pruning, a performance-oriented Approximate Computing technique: at inference time, the implementation randomly prunes (or skips) non-important weights (i.e., with low contribution to the \dnn{}'s accuracy) of the first layer, exponentially increasing the side-channel resilience of the protected \dnn{} implementation.
However, the original security analysis of \macpruning{} did not consider a control-flow dependency intrinsic to the countermeasure design. This dependency may allow an attacker to circumvent \macpruning{} and recover the weights important to the \dnn{}'s accuracy.
This paper describes a preprocessing methodology to exploit the above-mentioned control-flow dependency. Through practical experiments on a Chipwhisperer-Lite running a \macpruning{}-protected Multi-Layer Perceptron, we target the first $8$ weights of each neuron and recover $96\%$ of the important weights, demonstrating the drastic reduction in security of the protected implementation. Moreover, we show how microarchitectural leakage improves the effectiveness of our methodology, even allowing for the recovery of up to $100\%$ of the targeted non-important weights.
Lastly, by adapting our methodology, we elaborate on how the pruning mechanism, which depends on the importance of the weights, enables the circumvention of a control-flow-free \macpruning{} implementation. With this last point, we identify the pruning mechanism underlying \macpruning{} as the root of the countermeasure's vulnerability.
\end{abstract}

\section{Introduction}
\label{sec:introduction}
In a world where current challenges require the analysis and processing of large volumes of data, deep neural networks (\dnn{}s) are leading a fast and unexpected technological revolution.
From Convolutional Neural Networks to Generative Pre-Trained Transformers, \dnn{}s contribute to a multitude of fields, as automotive~\cite{singhDeepLearningAutomotive}, agriculture~\cite{AttriASR2023}, healthcare~\cite{miottoDeepLearningHealthcare2018} and cybersecurity~\cite{PicekPMWB2023}.
What makes \dnn{}s a powerful tool is their ability to automatically extract and learn, from a representative set of data for a specific problem, the required features to correctly work on new data. This learning process -- \emph{training} -- assigns to each neuron a set of \emph{weights}.
Weights represent a sensitive intellectual property (\ip{}) asset that product vendors or service providers want to keep confidential, mainly due to:
\begin{itemize}
    \item Expensive Training: training is a resource-intensive process that requires up to several weeks~\cite{costTimeTraining} and expensive hardware~\cite{costDollarTraining}, hence incurring high energy costs and contributing a significant impact on the carbon footprint~\cite{costEnergyCarbonTraining};
    \item Safety, Security, and Privacy Risks: knowledge of weight values may facilitate building other attacks, threatening the safety and privacy of end users or the security and privacy of a product/service (e.g., adversarial or model inversion attacks~\cite{CoqueCSZ2023, CoqueCSZ2024}). 
\end{itemize}

Widely used and studied in cryptography, passive side-channel analysis (\sca{}) has recently appeared as an effective approach to violate the secrecy of this \ip{}: through the statistical analysis of physical measurements (e.g., power consumption, EM emission, execution time) of the \dnn{} implementation, a malicious user can recover information on the weights.
Several works demonstrated the efficacy of \sca{} on different \dnn{} implementations in terms of architecture (Multi-Layer Perceptron, Convolutional Neural Network, Spiking Neural Networks), arithmetic (floating-point weights, integer weights), and running platform (\cpu{}, \gpu{}, \fpga{}) \cite{BatinBJP2019, JoudMPR2023, YuMYZJ2020, GaoQZWMAXFN2023, ProbsBS2025a, MajiBC2021, yli-mayryExtractionBinarizedNeural2021, ZhangDF2023, HorvaCWBY, thu2023bus, thu2023you, BatinBJP2021, HeWLZFL2022, mendez2021physical}.

State-of-the-art techniques to counteract \sca{} comprise \emph{masking}~\cite{DubeyCA2020a, DubeyCA2020, DubeyAPCA2022, MajiBFC2022, AthanWDF2022, BroscPGS2024} and \emph{hiding}~\cite{BroscPS2022, ZhangMHWHLLZYL2023, WuWZC2024} -- two classic side-channel countermeasures --, the application of multi-party computation~\cite{HasheRFG2022}, and \dnn{}-tailored solutions, as parasitic layers~\cite{ChabaDGK2022}.
These countermeasures trade off higher side-channel resilience with increased inference time, energy consumption, and resource utilisation.

Recently, Ding et al. have proposed a new countermeasure, \macpruning{}, inspired by pruning, a technique designed to improve the performance of \dnn{} implementations~\cite{DingGRDF25}: by skipping certain pixels with a given probability at inference time, the countermeasure desynchronises the side-channel traces and deprives the attacker of the required information to build the leakage hypotheses. As a result, the number of traces needed to
extract the weights through a non-profiled \sca{}, e.g., Correlation Power Analysis (\cpa{}), exponentially increases.
However, their implementation exhibits a control-flow dependency that may be exploited through pattern-matching techniques, nullifying the theoretical exponential security increase of the countermeasure.

In this paper, we evaluate the actual security of \macpruning{} in light of the control-flow-dependent nature of Ding et al.'s implementation. We contribute by:

\begin{itemize}
  \item Elaborating on how attackers can infer which pixels are processed or not through control-flow dependency (Section~\ref{sec:implementation-review});
  \item Describing a methodology to exploit the control-flow dependency in the implementation and circumvent the countermeasure (Section~\ref{sec:methodology});
  \item Experimentally validating the proposed methodology by recovering $96\%$ of the important weights from the first $8$ weights of each neuron of the input layer of a simple Multi-Layer Perceptron protected with \macpruning{} (Section~\ref{subsec:unprotected-implementation} and Section~\ref{subsec:protected-implementation});
  \item Demonstrating how an unintended information leakage -- potentially induced by the microarchitecture -- may improve our methodology, even allowing for the recovery of up to $100\%$ of the targeted non-important weights (Section~\ref{subsec:unintended-non-important-weights-recovery});
  \item Proving that, for any \dnn{} on which \macpruning{} can be applied (e.g., \mlp{}, \cnn{}), the pruning mechanism, which depends on the importance of the weights, is the actual root cause of the countermeasure vulnerability.
  To this end, we describe a control-flow-free \macpruning{} implementation and outline a modification to our methodology to circumvent the new countermeasure implementation (Section~\ref{sec:fix-macpruning}).
\end{itemize}

To ensure the reproducibility of our results and foster further research on \sca{} countermeasures for \dnn{}s, we open-source the implementation of our analysis tools and of the evaluated \macpruning{} implementation\footnote{\change{\repourl}}, the collected traces and the related inputs\footnote{\change{\dataseturl}}

\section{Background}
\label{sec:background}
\begin{figure}[t]
  \centering
  \includegraphics[scale=0.9]{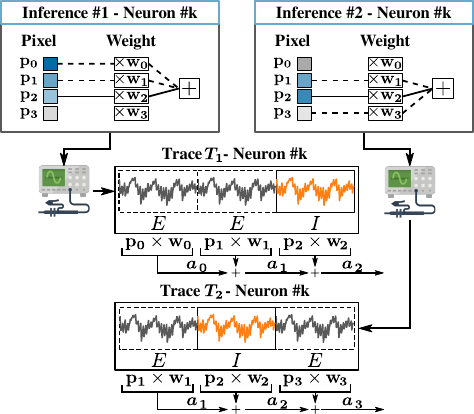}
  \caption{Application of \macpruning{} to the $k$-th neuron during two inferences and its effect on the side-channel traces.
           Solid lines represent important pixels, whereas dashed ones refer to executed non-important pixels.
           A missing path from pixel to the sum block indicates a skipped non-important pixel.}
           
  \label{fig:macpruning-example}
\end{figure}

This section provides the necessary background to understand the countermeasure and the principle underlying its \sca{} mitigation.

\subsection{Side-Channel-Driven DNN Weight Recovery}

\dnn{}s are network-like structures organised in connected \emph{layers}, each containing one or more nodes called \emph{neurons}.
The neuron represents the fundamental computation unit of a \dnn{}: from the given inputs, it combines them to synthesise a new input for the next layer.
We mathematically describe an $N$-input neuron as:

\begin{equation}
  y = f(\myvec{x})_{\myvec{w}} = \sigma(\Sigma_{i = 0}^{N - 1}(x_{i} \times w_{i})) + b
\end{equation}

where $\sigma(\cdot)$, $\myvec{x}$, $\myvec{w}, b$ represent, respectively, the activation function and the $N$-dimensional input, weight vector and bias.

Neurons' computations -- in particular, Multiply-and-Accumulate operations (\mac{}) -- leave a footprint (e.g., power consumption) statistically related to the processed inputs and intermediate values, which one can record in the form of a time-dependent signal $\mathbf{T}$ called \emph{side-channel trace}.

Under the condition the attacker has already recovered the weights $w_{j < i}$, and given $K$ possible values for a weight, they may recover $w_{i}$ by targeting the accumulation intermediate result $a_{i} = \Sigma_{j = 0}^{i} x_{j} \times w_{j}$ as follows~\cite{DingGRDF25}:

\begin{enumerate}
    \item generate a set of $N$ random inputs $x_{i}$;
    \item build the $(K, N)$ hypotheses matrix
        \begin{equation*}
            \left[ h_{k, i} = a_{i - 1} + w_{k} \times x_{i} \right];
        \end{equation*}
    \item measure $N$ traces $\mathbf{T}_{i}$ by recording the \mac{} activity when feeding the $x_{i}$ inputs to the target implementation;
    \item compute the $(K, N)$ score matrix
        \begin{equation*}
            \left[ s_{k, i} = d(\mathbf{T}_{i}, L(h_{k, i})) \right];
        \end{equation*}
    \item choose the $w_{k}$ that provides the best score $s_{k, i}$,
\end{enumerate}

where $d(\cdot, \cdot)$ is a statistical tool (e.g., Pearson's Correlation Coefficient) and $L(\cdot)$ is a \emph{leakage model} describing the side-channel leakage behaviour of the target implementation (e.g., the Hamming Weight).

\subsection{The MACPruning Countermeasure}
\macpruning{} relies on the concept of \emph{pruning}~\cite{hanDeepCompression2016}, a \dnn{} compression technique that removes unimportant computing elements of the network (e.g., \mac{}s, neurons) to deliver fast, lightweight, and energy-efficient implementations, at the cost of decreased accuracy.
Several works proposed some approaches to cope with this accuracy loss~\cite{HanLearnWeightsConns2015, ChengDNNPruning2024}.

Pruning is part of the Approximate Computing (\axc{}) paradigm~\cite{ArmenZSH2023}: a set of techniques that optimise computing systems' performance by approximating computations at different levels of the computing stack~\cite{mittalSurveyTechniquesApproximate2016, LeonHAJSPS2025b}.

\macpruning{} skips some of the \mac{}s running in the input layer, which processes the pixels of the input image.
To mitigate the negative impact on prediction accuracy, the countermeasure must distinguish pixels based on their \emph{degree of importance} in relation to their impact on overall accuracy.

Given this partition, the countermeasure skips non-important pixels with a user-defined probability $p$, leaving the processing (and order) of the important ones unchanged.
We refer to a \mac{} as \emph{(non-)important} if it processes a (non-)important pixel, and define IMAC and NIMAC as Important and Non-Important \mac{}; the same terminology applies to the weights.
For conciseness, we use $I, E$ and $S$ to respectively refer to important \mac{}s, non-important \emph{executed} \mac{}s, and non-important \emph{skipped} \mac{}s.

Skipping non-important \mac{}s with probability $p$ has an impact on the success of a weight recovery attack.
Firstly, the attacker analyses desynchronised traces, since the implementation skips different pixels with each inference: as exemplified by Figure~\ref{fig:macpruning-example}, during inference $\#1$, the implementation skips $p_{3}$, whereas, in the next inference, it skips pixel $p_{0}$. As a consequence, the implementation potentially uses the same weight (e.g., $w_{1}$) at different time instants.
Secondly, the accumulation intermediate $a_{i}$ depends on different weights with each inference: using the example in Figure~\ref{fig:macpruning-example}, $a_{2}$ depends on $w_{0},  w_{1}$ and $w_{2}$ in inference $\#1$, and on $w_{1}$ and $w_{2}$ in inference $\#2$. Therefore, the attacker has to foresee the skipped weights in each inference.
The combination of these phenomena exponentially increases the \emph{complexity} (i.e., the number of traces) to recover the weights of a network~\cite{DingGRDF25}.

\section{Exploiting the Control-Flow Dependency}
\label{sec:implementation-review}
This section dissects the \macpruning{} vulnerability that allows us to break its side-channel protection. We start describing the \macpruning{} threat model and continue providing a pseudo-code for the implementation described in the original paper, detailing each design step. Finally, we analyse this implementation and uncover how its control flow depends on the pixel importance, which, as we will show in Section~\ref{sec:methodology}, is at the core of the vulnerability. 
Through a practical example, built from traces measured using the Chipwhisperer-Lite platform hosting an \arm{} \cmfour{} (refer to Section~\ref{subsec:experimental-setup}), we show how such a dependency allows one to determine the pixel importance from the trace alone.
As the actual implementation from the original \macpruning{} paper was not released, we built our own version (pseudo-code in Section~\ref{subsec:macpruning-implementation}, assembler implementation in Section~\ref{subsec:control-flow-dependency}), whose full implementation source code we make openly available for the community (see Section~\ref{sec:introduction}) to reproduce, evaluate and build on our work.

\subsection{Threat Model}
\label{subsec:threat-model}
In this work, we adhere to the same threat model considered in the original \macpruning{} paper. We consider an attacker who aims to recover the weights used in a \dnn{} deployed on an edge device (e.g., a microcontroller).
The attacker has full knowledge of the \dnn{}'s hyperparameters (e.g., number of layers, activation functions, neurons per layer, which can be obtained through an architecture recovery attack~\cite{KuriaDYA2025} or if the \dnn{} model details are public) and of the \dnn{} implementation. They have no knowledge of the database used to train the \dnn{} implementation.
The attacker can only provide arbitrary inputs to the implementation, monitor its output, and measure (either physically or remotely) an analogue side channel (e.g., power consumption, EM emanation).
In addition, we consider that the attacker cannot eavesdrop on the transfer of weights and biases to the edge device.
Deployment frameworks and quantisation do not affect the applicability of our methodology.

\subsection{\macpruning{} Implementation}
\label{subsec:macpruning-implementation}
In the absence of a reference implementation, we used the high-level description from the original \macpruning{} paper to derive Algorithm~\ref{algo:inference-macpruning-pseudo-code}, which reports the pseudocode of a neuron's computation protected by \macpruning{}.
As part of the initialisation phase of the \dnn{} implementation, the end user sends the set of important pixels to the target device.
The implementation stores this set as a binary table of size $w \times h$ (with a value of 1 indicating an important pixel), with $w$ and $h$ the width and height of the input image, respectively.
As in the original formulation, we refer to this table as \emph{Importance-aware Pixel Activation Map} (IaPAM)~\cite{DingGRDF25}.
The transfer happens only once, and the table is left unmodified for all subsequent inference requests.

\begin{algorithm}
    \SetKw{from}{from}
    \SetKw{return}{return}
    \SetKwInOut{Input}{Input}
    \SetKwInOut{Output}{Output}
    \SetKwFunction{length}{length}

    \Input{$inputs$: $(w \times h)$-array of pixels\\
           $weights$: $(w \times h)$-array of weights\\
           $IaPAM$: $(w \times h)$-array of bits\\
           $randWords$: $(w \times h)$-array of bits}
    \Output{$acc$: accumulator}
    \BlankLine

    $acc \leftarrow 0$\;
    \For{$i$ \from $0$ \KwTo $\length(inputs)$}{\label{line:for-loop-check}
        \uIf{$IaPAM[i]$}{\label{line:check-iapam}
            $m \leftarrow inputs[i] \cdot weights[i]$\;\label{line:process-important-pixel-begin}
            $acc \leftarrow acc + m$\;\label{line:process-important-pixel-end}
        }
        \uElseIf{$randWords[i]$}{\label{line:check-to-execute}
            $m \leftarrow inputs[i] \cdot weights[i]$\;\label{line:process-non-important-pixel-begin}
            $acc \leftarrow acc + m$\;\label{line:process-non-important-pixel-end}
        }
        \uElse{
            \tcc{Skip MAC}\label{line:skip-mac}
        }
    }
  \return $acc$
  \caption{The pseudocode of a neuron's computation protected with \macpruning{}.}
  \label{algo:inference-macpruning-pseudo-code}
\end{algorithm}

First, the algorithm checks if the pixel is important (Line~\ref{line:check-iapam}).
If true, the neuron processes the pixel (Line~\ref{line:process-important-pixel-begin}--\ref{line:process-important-pixel-end}) and continues to the next one.
Otherwise, the neuron may skip the non-important pixel (Line~\ref{line:check-to-execute}) according to a randomly generated bit: if set to $1$, the neuron processes the pixel (Line~\ref{line:process-non-important-pixel-begin}--\ref{line:process-non-important-pixel-end}) and moves to the next one. Otherwise, the neuron does not accumulate the pixel contribution and steps to the next one.

\subsection{Control-Flow Dependency}
\label{subsec:control-flow-dependency}
In Algorithm~\ref{algo:inference-macpruning-pseudo-code}, determining whether a pixel is important (Line~\ref{line:check-iapam}) and whether to skip a non-important pixel (Line~\ref{line:check-to-execute}) involves control-flow statements, as extracted from \macpruning{} paper~\cite{DingGRDF25}, Section V.E.
These statements have a well-known impact on the implementation side-channel behaviour~\cite{Koche1996}: e.g., when running on a microcontroller, taking or not taking a branch may incur timing penalties~\cite{arm-m4-trm}. Furthermore, the implementation may perform different operations in the two branches.
As a consequence, a side-channel attacker may observe a difference in the length and shape of the trace related to the execution of the loop body (Line~\ref{line:check-iapam}--\ref{line:skip-mac}); this difference depends on the importance of the processed pixel.

\begin{lstlisting}[language={[ARM]Assembler}, basicstyle=\scriptsize\ttfamily, keywordstyle={\color{cerulean}\bfseries}, keywordstyle={[2]\color{pumpkin}\bfseries},keywordstyle={[3]\color{mortargrey}\bfseries}, frame=single, float, caption={Our \arm{} \thumbtwo{} \macpruning{} implementation}, label=lst:macpruning-implementation]
 <...>
 mov.w r0, #8
 .LoopExitCheck:
   cmp.w r0, #0
   cbz.n r0, .Exit
   sub.w r0, r0, #1

 .IMAC_check:
   ldr.w  r1, [pc, #264] @ Load IaPAM addr.
   ldrb.w r1, [r1, r4]
   mov.w  r2, #1
   lsl.w  r2, r2, r0
   ands.w r1, r1, r2     @ Is i-th pixel important?
   bne.w .MAC_computation
  
 .NIMAC_check:
   ldr.w  r1, [sp, #80] @ Load randomWords addr.
   ldrb.w r1, [r1, r4]
   mov.w  r2, #1
   lsl.w  r2, r2, r0
   ands.w r1, r1, r2    @ Execute i-th pixel?
   beq.w .LoopExitCheck
 
 .MAC_computation:
   ldr.w  r1, [sp, #76]  @ Load image addr.
   ldrb.w r1, [r1, r0]   @ Load pixel.
   ldr.w  r2, [sp, #84]  @ Load weights array addr.
   ldrb.w r2, [r2, r0]   @ Load weight.
   mla    r3, r2, r1, r3 @ Multiply-and-Accumulate.
   b.n .LoopExitCheck

 .Exit:
 <...>
\end{lstlisting}
To illustrate this phenomenon, we report our \macpruning{} implementation (in \arm{} \thumbtwo{}) in Listing~\ref{lst:macpruning-implementation}, which closely follows the logic described in Algorithm~\ref{algo:inference-macpruning-pseudo-code}:
the initial basic block \texttt{.LoopExitCheck} implements the loop check at Line~\ref{line:for-loop-check}. Basic blocks \texttt{.IMAC\_check} and \texttt{.NIMAC\_check} implement, respectively, the control-flow statements at Line~\ref{line:check-iapam} and Line~\ref{line:check-to-execute}, whereas basic block \texttt{.IMAC\_computation} implements the \mac{} computation (Line~\ref{line:process-important-pixel-begin}--\ref{line:process-important-pixel-end} and Line~\ref{line:process-non-important-pixel-begin}--\ref{line:process-non-important-pixel-end}).

\begin{figure}
    \centering
    \includegraphics[scale=0.8,keepaspectratio]{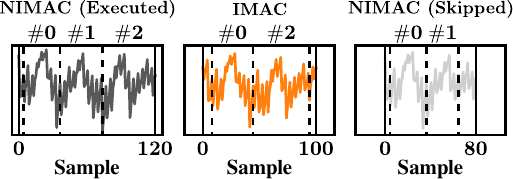}
    \caption{Side-channel patterns associated with important and non-important pixels identified from traces measured during the execution of Listing~\ref{lst:macpruning-implementation}.}
    \label{fig:cfg-effect}
\end{figure}

Figure~\ref{fig:cfg-effect} reports the unique side-channel patterns that we measure according to the importance of the pixel.
Each pattern exhibits a different length and shape. The three patterns share some subpatterns, which we delimit by dashed black lines: for instance, they start with subpattern $\#0$, which corresponds to the \texttt{.IMAC\_check} basic block; the implementation runs it to check the importance of a pixel.
We note that executing or skipping a non-importance pixel generates a different subpattern $\#1$: in the \emph{skipped} case, the branch is taken, completing in $1$ clock cycle; in the \emph{executed} case, the branch is not taken, requiring at least $2$ clock cycles to complete~\cite{arm-m4-trm}.

\begin{figure}
    \centering
    \includegraphics[scale=0.8]{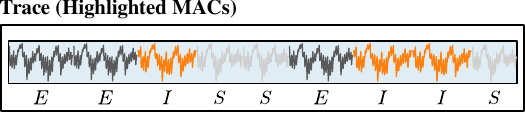}
    \caption{The application of the identified patterns to distinguish what \mac{} processes important and non-important pixels.
    We use $E$, $S$, and $I$ to indicate, respectively, non-important (executed and skipped) and important pixels.
    We measured the trace during the execution of Listing~\ref{lst:macpruning-implementation}.}
    \label{fig:macs-highlighted}
\end{figure}

As depicted in Figure~\ref{fig:macs-highlighted}, which shows a captured trace with the three different patterns (i.e., $E$, $S$ and $I$; respectively, non-important executed, non-important skipped, and important \mac{}s) highlighted in different colours, the control-flow dependency allows one to determine the important and the non-important (executed and skipped) pixels processed (dually, the \mac{}s and weights) along a given trace.

\section{Preprocessing Methodology}
\label{sec:methodology}
In the previous section, we showed how a side-channel attacker can observe the control-flow dependency from a trace and deduce the importance of the processed pixels.
Given this control-flow dependency, we propose a \emph{preprocessing} methodology to circumvent \macpruning{} and allow a side-channel attacker to recover the \emph{important weights} of the network. 
In brief, we captured $N$ traces from $N$ randomly generated images fed to the target \dnn{}, extracted the side-channel information on important weights, and fed it to the subsequent weight recovery phase.

\begin{figure}
  \centering
  \includegraphics[scale=0.9]{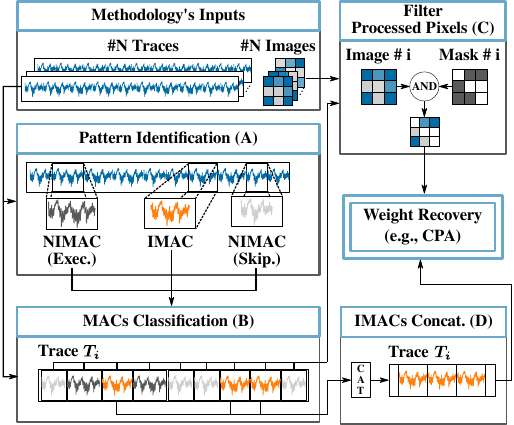}
  \caption{Proposed methodology to circumvent the \macpruning{} countermeasure. This methodology preprocesses the collected side-channel traces before the actual analysis phase (e.g., \cpa{}) and consists of $4$ steps: side-channel patterns identification (A), classification of \mac{}s operations (B), filtering out skipped pixels from input images (Step C) and concatenation of the side-channel patterns of important \mac{}s (Step D).
  }
  \label{fig:methodology}
\end{figure}

Figure~\ref{fig:methodology} illustrates the four steps of our methodology:
\begin{itemize}
    \item \textbf{Pattern Identification} (Section~\ref{subsec:pattern-identification}): the attacker uses its knowledge on the targeted platform and \dnn{} implementation to identify the side-channel patterns corresponding to important and non-important \mac{}s (IMAC and NIMAC, respectively);
    \item \textbf{\mac{}s Classification} (Section~\ref{subsec:macs-classification}): the attacker uses the identified patterns to assign a degree of importance to each \mac{};
    \item \textbf{Filter Processed Pixels} (Section~\ref{subsec:filter-processed-pixels}): the attacker uses the classification from the previous step to determine which pixel of the image the \dnn{} implementation processed; then, they filter out the unprocessed pixels from the original input images;
    \item \textbf{Important \mac{}s Concatenation} (Section~\ref{subsec:imacs-concat}): in parallel to the previous step, the attacker builds a new set of traces containing only the important \mac{}s, preserving their original order.
\end{itemize}
In the following paragraphs, we describe in detail each step of the preprocessing methodology.
We recall that $E$, $S$, and $I$ indicate, respectively, non-important executed, non-important skipped, and important \mac{}s.

\subsection{Pattern Identification}
\label{subsec:pattern-identification}
Mapping side-channel patterns to important and non-important pixels represents the first and most delicate step of our approach: a poorly-chosen set of patterns potentially prevents the correct classification of \mac{}s; without a correct classification, the methodology cannot evade the countermeasure. To minimise the risk of misclassification, the attacker should consider patterns that (1) do not overlap and (2) univocally map to a specific degree of importance (i.e., I, E, or S). Figure~\ref{fig:cfg-effect} reports an example of such patterns for the \macpruning{} implementation in Listing~\ref{lst:macpruning-implementation}.
According to our threat model (Section~\ref{subsec:threat-model}), one may use the knowledge on the \dnn{} implementation (i.e., assembler implementation) and target device to assign to each pattern a degree of importance (see Section~\ref{subsec:control-flow-dependency} for an example).
We note that, according to the device running the implementation, the same piece of code may exhibit different side-channel patterns, a common occurrence for microcontrollers with advanced microarchitectural features (e.g., multiple instruction issue). As a consequence, different patterns may map to the same degree of importance. Furthermore, the same piece of code, when computing different data, exhibits patterns that differ only in amplitude. For the sake of simplicity, we assume: (1) a one-to-one correspondence between patterns and degree of importance; (2) the amplitude difference does not hinder the correct classification of \mac{}s.

\subsection{MACs Classification}
\label{subsec:macs-classification}

\begin{algorithm}
    \small
    \SetKwInOut{Input}{Input}
    \SetKwInOut{Output}{Output}
    \SetKwFunction{lenght}{lenght}
    \SetKwFunction{dist}{dist}
    \SetKwFunction{argmax}{argmax}

    \Input{$T$: trace\\
           $P_{imac}$: Important MAC pattern\\
           $P_{exec}$: Non-Important MAC (Executed) pattern\\
           $P_{skip}$: Non-Important MAC (Skipped) pattern\\
           $threshold$: Pattern-matching confidence threshold}
    \Output{$P$: $Sample \mapsto Importance$}
    \BlankLine
    
    $l_{imac} \leftarrow \length(P_{imac})$\;
    $l_{exec} \leftarrow \length(P_{exec})$\;
    $l_{skip} \leftarrow \length(P_{skip})$\;

    $s \leftarrow T[0]$\;

    \While{$s \in T$}{
        $sim_{imac} \leftarrow \dist(T[s:l_{imac}], P_{imac})$\;\label{line:score_computation_begin}
        $sim_{exec} \leftarrow \dist(T[s:l_{exec}], P_{exec})$\;
        $sim_{skip} \leftarrow \dist(T[s:l_{skip}], P_{skip})$\;\label{line:score_computation_end}

        \tcp{List of (score, length, importance)}
        $scores \leftarrow [(sim_{imac},l_{imac}, I), (sim_{exec}, l_{exec}, E), (sim_{skip}, l_{skip}, S)]$\;\label{line:check_best_match_begin}
 
        \tcp{Get triple with highest score}
        $best\_match \leftarrow \argmax (scores)$\;
        \uIf{$best\_match.score \geq threshold$}{
            $s \leftarrow s + best\_match.length$\;
            $P[s] \leftarrow best\_match.importance$\;
        }\label{line:check_best_match_end}
        \Else{
            \tcp{No match, go to next sample}
            $s \leftarrow s + 1$\;\label{line:step_next_sample}
        }
    }
    \caption{Sliding-window Pattern Matching Strategy.}
    \label{algo:sliding-window-pattern-matching}
\end{algorithm}

From the identified patterns, the attacker can use any pattern-matching technique from the literature to classify the sequences of \mac{}s in the collected traces.
In this work, we use a sliding window strategy (Algorithm~\ref{algo:sliding-window-pattern-matching}).
Given a sample $s$ in the input trace $T$, we first compute a similarity score between each pattern $P_{x}$ obtained from step A and the piece of trace of length $l_{x}$ starting at sample $s$, with $x \in \{imac, exec, skip\}$ and $l_{x}$ the length of each of the three patterns (Line~\ref{line:score_computation_begin}--\ref{line:score_computation_end}).

We retain the pattern that reports the highest similarity score and that scores higher than a user-defined threshold. We assign to the analysed piece of trace the same importance (i.e., $E$, $S$, and $I$) as the retained pattern (Line~\ref{line:check_best_match_begin}--\ref{line:check_best_match_end}).
If none of the patterns score better than the threshold, we move on to the next sample in the trace (Line~\ref{line:step_next_sample}).
Finding a good threshold depends on the actual pattern-matching technique. In our case, we empirically experimented with different thresholds until we found one that minimised the classification error.

Figure~\ref{fig:methodology} reports an example of this sliding-window strategy on the $i$-th trace captured during the execution of a neuron.
Equation~\ref{eq:macs-classification-trace} reports the classification for this trace:
\begin{align}
    \textrm{Seq}_{i} &=\{S, E, I, E, S, S, I, I, S\}.\label{eq:macs-classification-trace}
\end{align}

The classification succeeds if, and only if, for all the derived sequences, the important \mac{}s are in the same position. Since the attacker does not have prior knowledge of the \iapam{}, they cannot check if the classification worked. To improve the reliability of the classification, they can partition the set of sequences according to the position of important pixels. The largest partition corresponds, with a certain probability, to sequences reporting a correct classification. This probability depends on the number of traces classified, their quality, and the quality of the patterns.
The attacker only preserves the correctly classified traces (and related input images) from the initial set of traces and images.

\subsection{Filter Processed Pixels}
\label{subsec:filter-processed-pixels}
In this step, the attacker preprocesses the input images to keep only the important and processed non-important pixels. Such a step is fundamental for the computation of the hypotheses required in the weight recovery phase.
Figure~\ref{fig:methodology} reports the masks and their application for the classification of a captured trace. Assuming a raster-scan pixel processing order, the attacker builds a mask with the same size as the images. A blank pixel in the mask means a skipped input pixel, whereas a grey one refers to a processed input pixel, important or not.
Then, they apply the corresponding mask to each image, so that the final image contains only processed pixels. The skipped pixels are set to an arbitrary value (e.g., 0). As a result, the attacker gets the images to use during the weight recovery phase.

\subsection{Important MACs Concatenation}
\label{subsec:imacs-concat}

In this final step, the attacker composes new side-channel traces from the previous classification step. Specifically, for each collected trace, they concatenate the parts of the trace related to important pixels. Since the attacker uses fixed-length patterns and the number of important pixels remains the same for each collected trace, the concatenation results in traces with the same length.
Furthermore, since the order of important pixels (thus of the \mac{}s) is the same in each trace, each concatenated piece relates to the same important pixel; hence, the concatenation results in vertically aligned traces.

\begin{algorithm}
    \small
    \SetKwInOut{Input}{Input}
    \SetKwInOut{Output}{Output}
    \SetKwFunction{cat}{cat}

    \Input{$T$: set of traces\\
           $P$: $T \times Sample \mapsto Importance$}
    \Output{$T_{R}$, set of traces (containing only important \mac{}s)}
    \BlankLine{}

    \For{$t \in T$}{
        $seq \leftarrow P[t]$\;\label{line:get_macs_sequence}
        $t_{R} \leftarrow [\;]$\;
        \For{$p \in seq$}{
            \If{$p.importance == I$}{\label{line:enqueue_important_patterns_begin}
                $t_{R} \leftarrow \cat(t_{R}, p)$\;
            }\label{line:enqueue_important_patterns_end}
        }
        $T_{R} \leftarrow t_{R}$\;\label{line:enqueue_new_trace}
    }
    
    \caption{Procedure to concatenate the Important MACs identified in step B.}
    \label{algo:realignment-algorithm}
\end{algorithm}

Algorithm~\ref{algo:realignment-algorithm} reports the realignment algorithm: for each trace $t$ in the input set $T$, we collect the extracted \mac{}s sequence (Line~\ref{line:get_macs_sequence}); then, for each entry $(position, importance)$ where $importance == I$ (i.e., the \mac{} is important), we concatenate the related pattern $p$ in the new trace $t_{R}$ (Line~\ref{line:enqueue_important_patterns_begin}--\ref{line:enqueue_important_patterns_end}).
Finally, we enqueue the new trace $t_{R}$ in the output set $T_{R}$ (Line~\ref{line:enqueue_new_trace}).

\section{Experimental Evaluation}
\label{sec:experimental-evaluation}
In this section, we describe our experimental software and hardware setup, the methodology we followed to analyse the security of \macpruning{}, and the results of our experiments.

\subsection{Experimental Setup}
\label{subsec:experimental-setup}
We ran our experiments on the ChipWhisperer-Lite side-channel platform (\cwlite{}), which hosts a \cmfour{} processor set to run at $7.37$ MHz.
We set the \cwlite{}' scope sampling rate to $4$ times the processor's frequency.

To evaluate our methodology, we considered a Multi-Layer Perceptron (\mlp{}), whose first layer -- protected by \macpruning{} -- consists of $5$ neurons, each fed with $32$-bit images.
We used TensorFlow (\texttt{v. 2.18.0}) to create the model, quantise it to 8 bits, and store it in \texttt{TFLite} format.
We used TinyEngine~\cite{tinyEngineRepo} (\texttt{commit 47bb283}) to generate a C version of the \texttt{TFLite} model.
We implemented \macpruning{} in \texttt{thumb-2} assembler (Listing~\ref{lst:macpruning-implementation}) and inlined it in the generated C code.
We compiled the C-based model via the \texttt{arm-none-eabi-gcc} compilation toolchain (\texttt{v. 15:10.3-2021.07-4}, optimisation level $3$).
We deployed the generated binary on the \cmfour{} processor.

\subsection{Security Analysis Methodology}

To assess the effectiveness of our preprocessing approach, we ran a \cpa{}-driven weight recovery on the \macpruning{}-protected \mlp{}.
Specifically, we targeted the weights $w_{1}$ to $w_{7}$ of each neuron in the first layer. We do not consider weight $w_{0}$ since its recovery depends on the attacker's ability to discern its true value among the possible candidates~\cite{BroscPS2022}.
We note that the recovery of a given weight $i$ depends on the successful recovery of all the previous weights $j < i$.
To better evaluate our method, we assumed that the attacker has already recovered all the previous weights $j$ when attacking weight $i$.
For each weight, we targeted the result of each \mac{} and used the Hamming Weight (\hw{}) leakage model to compute the leakage hypotheses.
In a side-channel attack, the true weight value may not coincide with the identified best weight value, but may appear among the first $n$ best values.
As such, the attacker has to verify which of the $n$ values is the true one.
The Guessing Entropy (\guessentr{}) quantifies in bits the average effort for a side-channel attacker to identify the true value~\cite{PapagGARRS2023}. In particular, a \guessentr{} value close to $0$ bits indicates that, on average, the true value scores the best.
To evaluate the recovery success of each weight, we report the \guessentr{} evolution with respect to the number of traces.
We computed the \guessentr{} on $5$ different sets of $50$k traces, collected by feeding the \mlp{} implementation with $50$k randomly generated 32-bit images.
In the following, we use the term \emph{experiment} to mean \emph{an evaluation carried out with one of these $5$ sets of random images}.

We considered three use cases: \mlp{} implementation with \macpruning{} disabled (unprotected implementation), with \macpruning{} enabled (protected implementation), and with \macpruning{} enabled and circumvented with our methodology.
To minimise discrepancies in results, we kept the integer values of the weights constant across all experiments, and used the same $5$ sets of random images that constitute each of the five experiments for each of the three use cases.

\subsection{Considering Extended Traces}
\label{subsec:consider-extended-traces}
\begin{figure*}
    \centering
    \includegraphics[scale=0.8,keepaspectratio]{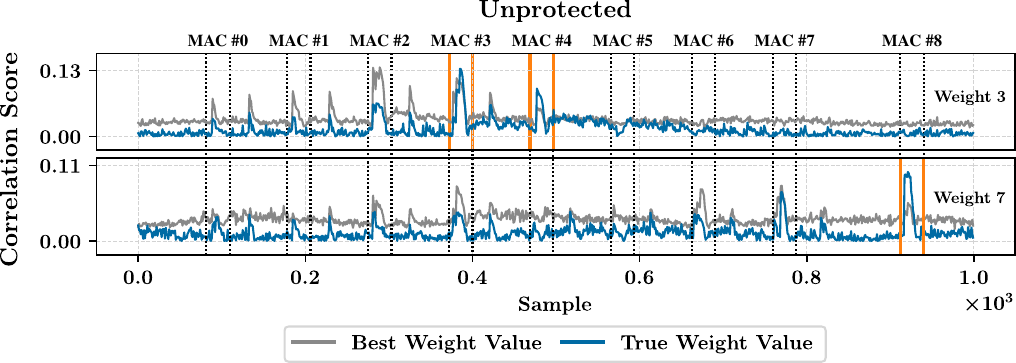}
    \caption{Mean correlation score (over the $5$ experiments, each consisting of $10$k traces) for weights $w_{3}$ and $w_{7}$ during the execution of the neuron $\#3$. We report the score for the true weight value and the best weight value.
    We highlight with solid orange lines the \mac{} where the true value scores the best.}
    \label{fig:unprotected-corrl-input-1-8-neuron-3}
\end{figure*}

During the analysis of the unprotected case, we observed that the implementation leaks information on weight $i$ during the execution of \mac{} $\#i$, \mac{} $\#i + 1$ or both.
Figure~\ref{fig:unprotected-corrl-input-1-8-neuron-3} illustrates this phenomenon in the unprotected case: the true value for weight $w_{3}$ shows the highest mean Pearson's Correlation score during \mac{} $\#3$ \emph{and} \mac{} $\#4$.
For weight $w_{7}$ the true value scores the best during \mac{} $\#8$, but not during \mac{} $\#7$.
Several works show that the microarchitecture of the \cmfour{} lies at the origin of unintended information leakage~\cite{BarenBIP2021, MarshPW2022, CasalBCH2023}. However, these works limited their analyses to instructions common to implementations of symmetric cryptosystems.
We conjecture that the particular implementation of the \texttt{mla} \thumbtwo{} instruction, whose leakage behaviour has not been studied before, may lie at the origin of the observed phenomenon.
To not lose the information leakage on weight $w_{7}$, which may be captured by \mac{} $\#8$, we analyse the traces to consider also the execution of this last one.

\subsection{Unprotected Implementation Results}
\label{subsec:unprotected-implementation}
\begin{figure}
    \centering
    \includegraphics[width=0.9\columnwidth,keepaspectratio]{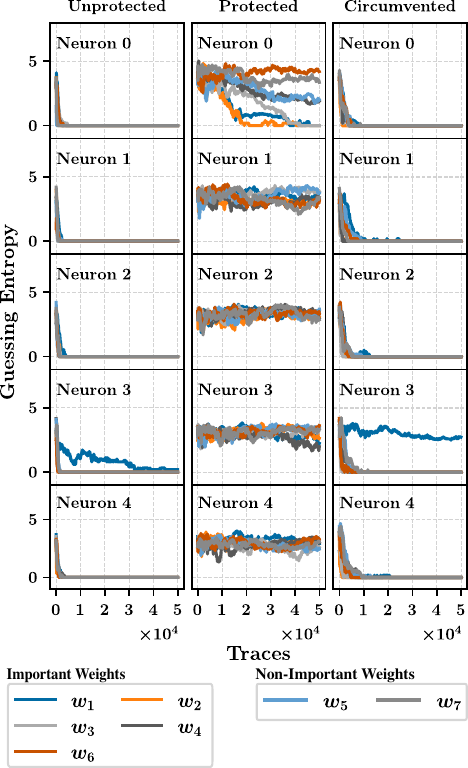}
    \caption{Guessing Entropy (GE) vs. number of traces for weights $w_{1}$ to $w_{7}$ of each neuron with \macpruning{} disabled (left), enabled (middle), and circumvented (right).}
    \label{fig:ge-input-1-8}
\end{figure}

Figure~\ref{fig:ge-input-1-8} (left column) reports the \guessentr{} for the unprotected implementation. The metric rapidly converges to $0$, indicating a successful recovery for almost all weights, except weight $w_{1}$ during the execution of neuron $\#3$: the \guessentr{} slowly decreases to $0.2$ within $50$k traces.
As the \guessentr{} is an averaged metric, we conclude that, for certain experiments, the true weight value does not score the highest. We conjecture that the particular set of input images may contribute to this phenomenon, although other causes (e.g., the microarchitecture) may also contribute.
In conclusion, the standard \cpa{}, carried on the unprotected \mlp{} implementation, recovers $97.14\%$ of the true weight values (i.e., 34 out of 35) in less than $50$k traces.

\subsection{Protected Implementation Results}
\label{subsec:protected-implementation}

For the protected implementation, we randomly generated the \iapam{}, loaded it to the \cmfour{} before each of the $5$ experiments, and kept it fixed through each inference.
Equation~\ref{eq:iapam} reports the \iapam{} bits for weights $w_{0}$ to $w_{7}$, which are the target of our analysis. A set bit means an important weight, whereas an unset bit means a non-important one.

\begin{equation}
    \text{\iapam{}} = \{\underset{\substack{\uparrow\\w_{0}}}{0},
        \underset{\substack{\uparrow\\w_{1}}}{1},
        \underset{\substack{\uparrow\\w_{2}}}{1},
        \underset{\substack{\uparrow\\w_{3}}}{1},
        \underset{\substack{\uparrow\\w_{4}}}{1},
        \underset{\substack{\uparrow\\w_{5}}}{0},
        \underset{\substack{\uparrow\\w_{6}}}{1},
        \underset{\substack{\uparrow\\w_{7}}}{0},
        \underset{\substack{\uparrow\\w_{8}}}{1}\}.
    \label{eq:iapam}
\end{equation}

We instantiated the countermeasure with an activation ratio $p = 0.5$ (\emph{i.e.}, skip non-important pixels with a $50\%$ probability) for each pixel.
We argue that such a choice represents the worst-case scenario for an attacker.
With $p \neq 0.5$, the implementation skips more or fewer non-important pixels, reducing the desynchronisation effect. Therefore, it becomes easier to identify the position of important \mac{}s.

We randomly generated $50$k binary tables of $32 \times 5$ entries (one entry for each weight of the \mlp{}'s first layer) and loaded a new table to the \cmfour{} before each inference. The implementation accessed the $i$-th entry of this table to check whether to execute or skip the $i$-th non-important \mac{}. For each of the $5$ experiments, we used the same set of $50$k tables.

We firstly present the results and their analysis without circumvention of \macpruning{}; we then report the results with circumvention of the countermeasure by application of the preprocessing methodology proposed in this paper.
We note that we used the same set of traces in both cases where \macpruning was activated, i.e., we did not repeat the experiment.

\noindent\textbf{\emph{Without circumvention:}}
Figure~\ref{fig:ge-input-1-8} (middle column) reports the evolution of the \guessentr{} metric for the protected case.
For all the neurons except the first, \macpruning{} prevents the convergence of the \guessentr{} to $0$; the countermeasure prevents weight recovery with $50$k traces.
Concerning neuron $\#0$, the \guessentr{} for weight $w_{1}, w_{2}$ and $w_{3}$ converges to $0$, although slower than in the unprotected case.
We justify this observation by noting that weights $w_{1}, w_{2}$ and $w_{3}$ are important weights; the traces always carry information on them. However, weight $w_{0}$, a non-important one, is skipped with probability $0.5$, increasing the number of traces required to recover the next weights.
Therefore, although the desynchronisation induced by randomly skipping weight $w_{0}$, $50$k traces are enough to recover the next $3$ important weights.
We further observe that the \guessentr{} for weight $w_{4}$ did not converge to $0$, although being an important weight. We find a potential explanation in the phenomenon reported in Section~\ref{subsec:consider-extended-traces}: the implementation leaks information on weight $4$ only during the non-important \mac{} $\#5$; the random skipping of \mac{} $\#0$ and \mac{} $\#5$ prevented the recovery of weight $w_{4}$ with $50$k traces.
In conclusion, through standard \cpa{}, we recover the true values of weights $w_{1}, w_{2}$ and $w_{3}$ with $50$k traces; that is, $12\%$ of the important weights (i.e., 3 out of 25), showing the effectiveness of the countermeasure, as expected.

\begin{figure}
    \centering
    \includegraphics[width=0.7\columnwidth,keepaspectratio]{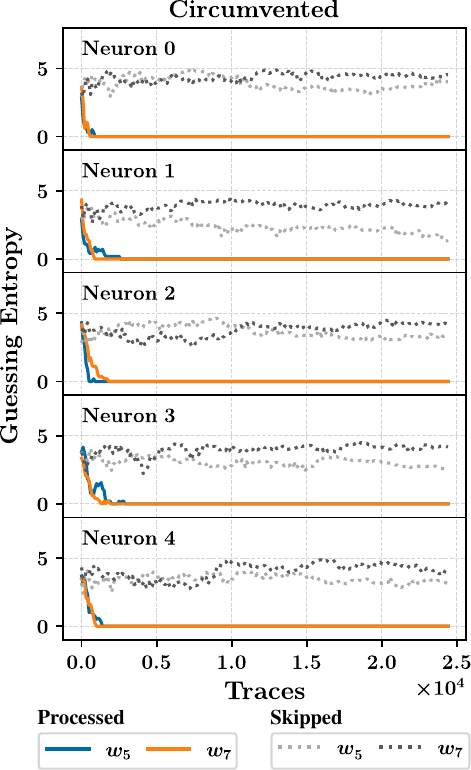}
    \caption{Comparison of \guessentr{} vs. number of traces for weight $w_{5}$ and $w_{7}$ when  processed and when skipped. We partitioned the preprocessed traces into two sets, depending on whether the two weights are processed or not.}
    \label{fig:circumvented-ge-input-5-7-filtered}
\end{figure}

\noindent\textbf{\emph{With circumvention:}}
Figure~\ref{fig:ge-input-1-8} (right column) reports the evolution of the \guessentr{} metric when preprocessing the traces with our approach.
The \guessentr{} of the important weights $w_{1}, w_{2}, w_{3}, w_{4}$ and $w_{6}$ quickly converge to $0$, indicating the successful recovery of their true value in each of the $5$ experiments.
For $w_{1}$ of neuron $\#3$, the metric does not converge. This result is consistent with the one for the unprotected case, where $w_{1}$ does not converge with $50$k traces.
As previously discussed for the unprotected case, we conjecture that the used set of input images may explain the result, although other causes may also contribute to it.
In conclusion, our methodology allowed for the recovery of $96\%$ of the important weights (i.e., 24 out of 25), with less than $50$k traces. Due to similar results with the unprotected case, we conjecture that the missed recovery of $w_{1}$ is not attributable to our approach; potentially, our technique may allow one to recover all the important weights of the first layer, defeating \macpruning{} with a number of traces comparable to an unprotected implementation.
Counter-intuitively, the \guessentr{} of non-important weights $w_{5}$ and $w_{7}$ converge to $0$ too.
In the next section, we analyse this result.

\subsection{Unintended Recovery of Non-important Weights}
\label{subsec:unintended-non-important-weights-recovery}
Through our \cpa{} on the preprocessed traces, we were also able to recover the true values for $w_{5}$ and $w_{7}$.
This means that the preprocessed traces carry information on non-important weights. 
This is an unexpected result, as we designed our methodology to preserve parts of the traces where important weights are processed.
We account the phenomenon described in Section~\ref{subsec:consider-extended-traces} for this result: the implementation leaks information on weight $w_{5}$ during the processing of weight $w_{6}$, an important weight; the same observation holds for weight $w_{7}$, which leaks while processing the important weight $w_{8}$. Thus, when a non-important \mac{} is immediately followed by an important one, the preprocessed traces carry information on the former as well.
We verified this hypothesis for weights $w_{5}$ and $w_{7}$ by partitioning the traces in a \emph{processed set}  (weights always processed) and in a \emph{skipped set} (weights always skipped).
Due to the probabilistic nature of the countermeasure, the two sets do not have exactly the same number of traces. To avoid any discrepancy due to a mismatch in the number of traces, we analysed the first $24.5$k traces in both sets.

Figure~\ref{fig:circumvented-ge-input-5-7-filtered} reports the \guessentr{} for weights $w_{5}$ and $w_{7}$: for the skipped set (dotted line), \guessentr{} does not converge to $0$ with $50$k traces, whereas for the processed set (solid line), \guessentr{} converges to $0$, confirming our hypothesis.
This unintended information leakage enhanced our methodology, allowing for the recovery of $100\%$ of the targeted non-important weights.

We generalise this result to consider the case where $k > 0$ non-important weights separate a non-important weight $w_{j}$ from the next important weight $w_{i}$:

\begin{equation*}
    w_{j} \rightarrow \underbrace{w_{j + 1} \rightarrow \ldots \rightarrow w_{i -1}}_{k\ \text{non-important weights}}\rightarrow w_{i}.
\end{equation*}

We prove that the number of traces to recover $w_{j}$ through the analysis of $w_{i}$ exponentially increases with $k$.

\begin{proof}
    The implementation skips any of the $k$ weights with probability $p = 0.5$. 
    Thus, $w_{i}$ immediately follows $w_{j}$ (i.e., all the $k$ weights are skipped) with a rate $r = \prod_{h = 0}^{k - 1} p = \frac{1}{2^{k}}$.
    If we require $N$ traces to recover the true value of $w_j$ when $k = 0$, for $k \geq 1$ we would require $N' = \frac{N}{r} = N \cdot 2^{k}$ traces.
    That is, the required number of traces increases exponentially with the number of non-important weights $k$ between the targeted $w_{j}$ and the important weight $w_{i}$.
\end{proof}

As shown, this unintended information leakage enhances our preprocessing technique, allowing for the recovery of certain non-important weights.
In conclusion, secure side-channel countermeasures require careful implementation choices (e.g., control-flow statements) and deep knowledge of leakage behaviour on the target platform (e.g., \cpu{}s microarchitecture).

\section{Can We Fix \macpruning{}?}
\label{sec:fix-macpruning}
Our methodology circumvents \macpruning{} by taking advantage of the control-flow dependency in Ding et al.'s implementation, which exposes distinct side-channel patterns for important, non-important executed and non-important skipped \mac{}s.
However, we argue that the pruning mechanism, conditioned on the importance of the weights, is the root cause enabling the evasion of the countermeasure.
To prove this, we first describe an enhanced control-flow-free \macpruning{} inference. 
Then, we show how even in the presence of this hardened variant, a modification in our methodology re-enables successful recovery of important weights.

\begin{algorithm}
    \small
    \SetKw{from}{from}
    \SetKw{return}{return}
    \SetKwInOut{Input}{Input}
    \SetKwInOut{Output}{Output}
    \SetKwFunction{length}{length}
    \SetKwFunction{toSkip}{toSkip}
    \SetKwFunction{skip}{skip}
    \SetKwFunction{exec}{exec}
    \SetKwFunction{call}{call}

    \Input{$inputs$: $(w \times h)$-array of pixels\\
           $weights$: $(w \times h)$-array of weights\\
           $IaPAM$: $(w \times h)$-array of bits\\
           $randWords$: $(w \times h)$-array of bits}
           
    \Output{$acc$: accumulator}
    \BlankLine

    $acc \leftarrow 0$\;
    \For{$i$ \from $0$ \KwTo $\length(inputs)$}{
        $cond \leftarrow \toSkip(IaPAM[i], randWords[i])$\;\label{line:toSkip}
        $fn \leftarrow cond * addr(\exec()) + (1 - cond) * addr(\skip())$\;\label{line:choose-fn}
        $acc \leftarrow fn(inputs[i], weights[i], acc)$\;\label{line:call-fn}
    }
    \return $acc$\;
  \caption{Pseudo-code of a generic inference procedure enhanced with a control-flow-free \macpruning{} implementation.}
  \label{algo:inference-macpruning-branchless-pseudo-code}
\end{algorithm}

Algorithm~\ref{algo:inference-macpruning-branchless-pseudo-code} reports our \macpruning{} control-flow-free version: a side-channel secure function \func{toSkip} checks whether to skip a \mac{} (Line~\ref{line:toSkip}), retrieves the address of the function to call (\func{exec} or \func{skip}) (Line~\ref{line:choose-fn}), and executes it (Line~\ref{line:call-fn}).
The function \func{exec} executes the \mac{} and returns the result, whereas \func{skip} returns the accumulator value $acc$ given as input.
By design, \func{exec} and \func{skip} implementations differ; thus, their executions show distinct side-channel patterns.
Therefore, an attacker may still understand whether a non-important \mac{} is skipped, but not if a \mac{} processes an important pixel: Algorithm~\ref{algo:inference-macpruning-branchless-pseudo-code} uses \func{exec} for both.
Yet, the pattern difference between executed and skipped \mac{}s is enough to evade the enhanced countermeasure with a slightly modified version of our methodology.
In the \mac{}s Classification step (Section~\ref{subsec:macs-classification}), the attacker would use the results of the classification to derive the \iapam{}: they mark as important a \mac{} $i$ if, and only if, the classification reports it as executed across all traces. To exemplify, let us consider two \mac{}s sequences obtained by the classification step:

\begin{align}
    \textrm{Seq}_{1} &=\{S, E,
    \boldsymbol{E},
    E, S, S,
    \boldsymbol{E},
    \boldsymbol{E},
    S\}\label{eq:macs-classification-trace-1},\\
    \textrm{Seq}_{2} &=\{E, E,
    \boldsymbol{E},
    S, S, E,
    \boldsymbol{E},
    \boldsymbol{E},
    S\}\label{eq:macs-classification-trace-2}.
\end{align}

We can only distinguish whether a \mac{} is skipped, hence the classification does not report what the important \mac{s} are; we indicate them in bold in the sequences.
By preserving the positions where there are only executed \mac{}s, we have $\text{\iapam{}} = \{0, 1, 1, 0, 0, 0, 1, 1, 0\}$.
We notice that the derived \iapam{} marks the \mac{} in position $1$ as important, although it is not.
Thus, a correct \iapam{} recovery depends on the number of traces classified.
Given the \iapam{}, it is possible to extract the portions of the trace capturing the execution of important \mac{}s and use them for the concatenation step of the methodology (Section~\ref{subsec:imacs-concat}).

Therefore, the execution of different operations (i.e., \func{exec} and \func{skip}) according to the pixel importance allows for the recovery of important weights, with a certain probability. This probability depends on the number of traces required to correctly recover the \iapam{} provided to the \dnn{} implementation.
We further remark that whatever model on which \macpruning{} is applicable (e.g., \mlp{}, \cnn{}) will suffer from the same weakness.

In conclusion, we have proved that by leveraging pruning -- the very same mechanism on which \macpruning{} relies to deliver side-channel security -- an attacker can practically evade the countermeasure.

\section{Related Work}
\label{sec:related-work}
In this section, we survey existing countermeasures against weight-recovery attacks and pattern-matching techniques to evade side-channel countermeasures.

\subsection{Side-channel Defences against Weight Recovery Attacks}
Most of the side-channel countermeasures for DNNs rely on masking and hiding, two widely approaches issued from the much more mature field of cryptographic implementations hardening.

\emph{Masking} randomises the data on which implementations work, breaking the statistical link between the measured side channel and the target secret information, enabling provable security at the cost of a quadratic increase in implementation overhead.
Different works have addressed the challenge of applying masking on \dnn{} implementations while limiting the impact on performance/area overheads~\cite{DubeyCA2020a, DubeyCA2020, DubeyAPCA2022, AthanWDF2022, MajiBFC2022, BroscPGS2024}.
\emph{Hiding}, instead, buries secret-dependent signals under noise using techniques like:
\emph{shuffling} the execution order of \mac{} operations and neurons~\cite{BroscPS2022};
switching the frequency and clock phase of \mac{} hardware among pareto-optimal points~\cite{ZhangMHWHLLZYL2023};
using \emph{dual-rail precharged logic} to keep power consumption constant along inference, whatever data the \dnn{} computes~\cite{WuWZC2024}.

Other techniques include:
multi-party computation, which can also help build a side-channel secure \dnn{} inference engine~\cite{HasheRFG2022};
and adding a small Convolutional Neural Network (CNN) to the input layer of the target \dnn{} to approximate its inputs and complicate weight recovery~\cite{ChabaDGK2022}.

All of these countermeasures increase the implementation's overhead (execution time, energy consumption, resource utilisation).
Inspired by \axc{}~\cite{ArmenZSH2023,mittalSurveyTechniquesApproximate2016,LeonHAJSPS2025b}, \macpruning{}~\cite{DingGRDF25} introduced the first approximation-based side-channel countermeasure.
Through pruning, the countermeasure desynchronises the traces and deprives the attacker of the information required to successfully recover the weights.
Beyond side-channel vulnerabilities,
\axc{} is progressively finding a wider use to protect \dnn{}s against a large variety of privacy-oriented attacks, such as Adversarial Attacks~\cite{DBLP:conf/iolts/KhalidATHRA019, DBLP:conf/asplos/GuesmiAKBFAA21} and Model Inversion Attacks~\cite{IslamOAK2023}.

\subsection{Pattern-matching-based Preprocessing}
Our methodology -- the first one to deal with \macpruning{} -- uses pattern matching to resynchronise the side-channel traces, and to cope with the deprivation of the information required to run a successful weight recovery attack.
To the best of our knowledge, our work is the first one to address a countermeasure relying on information deprivation.
However, other works propose the use of pattern matching to circumvent countermeasures that desynchronise the traces through the insertion of random delays.
We partition these works into random delays removal, and attack points preservation.
The \emph{random delays removal} approaches identify the side-channel patterns corresponding to the random delays and using pattern matching to remove them from the traces~\cite{DurvauxRSOV12, StrobelP12, tianGeneralApproachPower2012}.
\emph{Attack points preservation} relies on pattern matching to identify, along the measured traces, points of interest for the side-channel attack (e.g., memory accesses to \texttt{sbox}es). To resynchronise the traces, the attacker only preserves the identified points of interest~\cite{AbdelCPJ17, TianSSH2012}. 

Our approach is closer to the second category, as it resynchronises the traces by preserving and vertically aligning the trace samples corresponding to the execution of the important \mac{}s (i.e., our attack points).
Yet, to counteract \macpruning{}, we also need to nullify the information deprivation. For this, we apply a strategy similar to the first category: we identify and spot the non-important \mac{}s (i.e., the points responsible for the desynchronisation), along the traces.

\section{Conclusion}
\label{sec:conclusion}
 In this paper, we described a practical methodology to circumvent \macpruning{}, a \dnn{}-oriented side-channel countermeasure relying on pruning, a performance-oriented \axc{} technique.
 \macpruning{} acts on the input layer, and it skips non-important input pixels -- and the corresponding weights -- to exponentially increase the security of the protected \dnn{} implementation.
 Our methodology takes advantage of the side-channel footprint left by pruning to identify and extract, at the cost of a simple non-profiled vertical attack (e.g., \cpa{}), the important weights, evading the countermeasure.
 We experimentally validated our methodology on a protected implementation of an \mlp{} running on a \cmfour{} microcontroller hosted on the Chipwhisperer Lite side-channel evaluation platform. We targeted the first $8$ weights of each neuron, and we recovered $96\%$ of the important weights.
 Furthermore, we showed how an unintended information leakage -- potentially attributable to the \cmfour{}'s microarchitecture -- improves our methodology, allowing for the recovery of up to $100\%$ of the targeted non-important weights.
 As a result, either implemented through control-flow statements or not, we have also shown how pruning itself leaks information that a side-channel attacker, independently of the protected \dnn{} (e.g., \mlp{}, \cnn{}) can leverage to evade the countermeasure.
 Future works will explore side-channel-aware designs for \macpruning{} and, more in general, the opportunities and risks of using \axc{} to build side-channel countermeasures.

\section*{Acknowledgements}
Work funded by the French \emph{Agence Nationale de la Recherche} (ANR) Young Researchers (JCJC) program, grant number ANR-21-CE39-0018 project ATTILA and ANR-23-E39-0003-01 project CoPhyTEE.

\bibliographystyle{IEEEtran}
\bibliography{biblio}

\end{document}